\documentstyle[epsf,aps,twocolumn,prl,floats]{revtex}

\def\mydate{1 August 2000}
\def\ignore#1{{}}

\newcommand{\beeq}{\begin{equation}}
\newcommand{\eneq}{\end{equation}}
\newcommand{\beqn}{\begin{eqnarray}}
\newcommand{\eeqn}{\end{eqnarray}}

\def\la{\raise.16ex\hbox{$\langle$}\lower.16ex\hbox{}  }
\def\ra{\, \raise.16ex\hbox{$\rangle$}\lower.16ex\hbox{} }
\def\go{\rightarrow}

\def\psibar{ \psi \kern-.65em\raise.6em\hbox{$-$} \lower.6em\hbox{} }

\title{\bf False Vacuum Black Holes and Universes}

\author{{\bf R.\ G.\ Daghigh   and \bf J.\ I.\ Kapusta}\\
 {\small \it School of Physics and Astronomy, University of
Minnesota, Minneapolis, MN 55455}\\
\vspace*{0.1in}
{\bf Y.\ Hosotani}\\
{\small \it Department of Physics,  Osaka University,
Toyonaka, Osaka 560-0043, Japan}\\
}

\date{\mydate}

\begin{document}

\draft

\wideabs{

\rightline{\small NUC-MINN-00/12-T}
\rightline{\small OU-HET 354}
\rightline{\small UMN-TH-1916/00}

\vskip .5cm
\maketitle

\begin{abstract}
We construct a black hole whose interior is the false vacuum and whose
exterior is the true vacuum of a classical field theory.  From the outside
the metric is the usual Schwarzschild one, but from the inside the space is
de Sitter with a cosmological constant determined by the energy of the
false vacuum.  \ignore{The energy of this configuration is finite.}  The
parameters of the field potential may allow for the false vacuum to exist
for more than the present age of the universe.  A potentially relevant
effective field theory within the context of QCD results in a
Schwarzschild radius of about 200 km.
\end{abstract}


}

Normally one thinks of black holes as being created in the collapse of a
star which was originally 10 to 20 times the solar mass.  It is
also believed that there are huge black holes of about 1 million times
the solar mass near the center of galaxies.  Another possibility
are primordial black holes, which were created in the early universe, which
presently have a mass less than 1\% that of the earth, and are microscopic
in dimension.  One does not usually ask what goes on inside a black hole
because there is an event horizon \ignore{at the Schwarzschild radius}
which prevents information from leaving it.  In this paper we study a black
hole whose interior is completely filled by the false vacuum of a classical
field theory.

The type of potential we have in mind is drawn in Figure 1. For the present
purpose it may conveniently be represented as a fourth-order polynomial in
the real classical field $\phi$:
\begin{eqnarray}
V(\phi)&=& \frac{\lambda}{4}(\phi-f_+) \nonumber \\
&& \hskip -.5cm  \times  \bigg\{ \phi^3 + \frac{4f_--f_+}{3}
\phi^2 -\frac{2f_-+f_+}{3} f_+ (\phi+f_+) \bigg\} \nonumber \\
V^{\prime}(\phi)&=& \lambda \phi(\phi+f_-)(\phi-f_+) \, .
\end{eqnarray}
Here $f_+ > f_- > 0$.  Defining $f=(f_++f_-)/2$ and $\Delta f = f_+-f_-$
the false vacuum at $\phi=-f_-$ has energy density $\epsilon_- = V(-f_-) =
\frac{2}{3}\lambda f^3 \Delta f$ while the true vacuum at $\phi=f_+$ has
zero energy density.

We look for time independent, spherically symmetric solutions of
Einstein's and Lagrange's equations with the metric
\begin{equation}
d\tau^2 = \frac{H(r)}{p^2(r)}dt^2 - \frac{dr^2}{H(r)} -r^2 d\Omega^2 \, .
\end{equation}
The action is
\begin{equation}
I = \int d^4x \sqrt{-g} \left[\frac{1}{16\pi G}R +\frac{1}{2}g^{\mu\nu}
\partial_{\mu}\phi\partial_{\nu}\phi - V(\phi) \right] \, .
\end{equation}
Defining $H(r) = 1 - 2GM(r)/r$, the equations of motion are
\begin{eqnarray}
&&\frac{1}{p}\frac{dp}{dr} = -4\pi G r \left(\frac{d\phi}{dr}\right)^2 
\label{Ein1} \\
&&\frac{dM}{dr} = 4\pi r^2 \left[ \frac{1}{2}H
\left(\frac{d\phi}{dr}\right)^2 + V(\phi)\right] \label{Ein2} \\
&&\frac{p}{r^2} \frac{d}{dr}\left( r^2 \frac{H}{p} \frac{d\phi}{dr}\right)
= V^{\prime}(\phi) \,  \label{Ein3} .
\end{eqnarray}
A nontrivial solution is $\phi = -f_-$ for $r < r_0$ and  $\phi = f_+$ for
$r > r_0$ where $r_0$ is arbitrary.  The discontinuity in the field at
$r_0$ will be analyzed below for the case of most interest to us.

\begin{figure}[bth]\centering
\leavevmode 
\mbox{
\epsfxsize=8.0cm \epsfbox{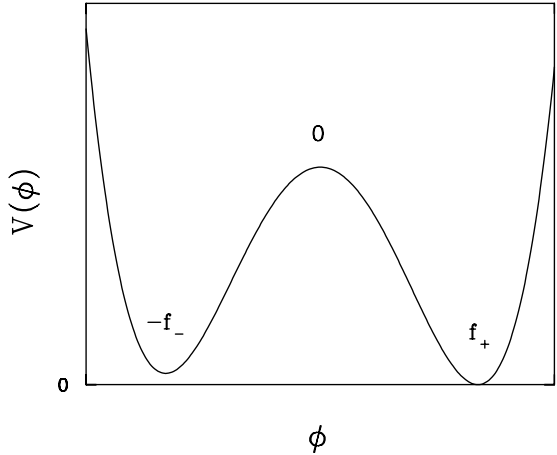}}
\caption{
An example of a scalar potential with one local
minimum (false vacuum) and one global minimum (true minimum).  The
potential drawn is a fourth-order polynomial.}
\label{fig_pot}
\end{figure}

For the above solution the metric functions are given by
\begin{eqnarray}
p&=& p_{\rm in}={\rm constant} \nonumber \\
H&=&1-\frac{r^2}{r_c^2}
\end{eqnarray}
when $r<r_0$ and
\begin{eqnarray}
p&=& p_{\rm out}={\rm constant} \nonumber \\
H&=& 1-\frac{r_0^3}{r_c^2 r}
\end{eqnarray}
when $r>r_0$.  The two constants $p_{\rm in}$ and $p_{\rm out}$ are not 
the same; generally one of them may be set to one by scaling the time
variable.  The $r_c$ is a critical radius related to a cosmological
constant $\Lambda=3/r_c^2$ and given by
\begin{equation}
r_c = \sqrt{\frac{3}{8\pi G \epsilon_-}} =
\frac{3m_P}{4f\sqrt{\pi \lambda f \Delta f}}
\end{equation}
Here $ m_P =G^{-1/2}  = 1.22\times 10^{28}$ eV is the Planck mass.  The
function $H(r)$ is plotted in Figure 2.  It has a cusp at the edge of the
bubble, $r=r_0$.    For $r_0<r_c$ we have a  false vacuum bubble,
which may be unstable,  in a space with zero cosmological constant and with
the gravitational field taken into account.  When $r_0=r_c$
there is a single horizon.  From the outside it appears as a black hole with
Schwarzschild radius
$r_S = r_c$.  The outside metric is just the usual Schwarzschild one as
guaranteed by the Birkhoff Theorem.  When
$r_0>r_c$, $H$ crosses zero twice and there are two horizons, an inner one
at $r_c$ and an outer one at $r_S = r_0^3/r_c^2 > r_c$.  Anyone on the
outside would only see the Schwarzschild solution, a black hole, and would
have no idea what is inside.  For this last case $g_{tt}<0$ and $g_{rr}>0$
in the shell $r_c<r<r_S$, indicating that the time and radial coordinates
have switched roles.   In this shell one can make a change of coordinates,
redefining time, after which the solution we have found is no longer time
independent in the new variables.

The case of interest in this paper is when $r_0=r_c$, that is, a black
hole with false vacuum on the inside and true vacuum on the outside. 
The fact that no information can cross the boundary is recognized
in two different manners.  For an outside observer no signal from inside
can cross the event horizon of the Schwarzschild black hole.  For an
inside observer any signal emitted outwards can reach at most the
cosmological horizon as the de Sitter space expands.  Though  completely
different, these describe the same physical event.

The first task is to study the behavior of $\phi$ near the surface.  Is
the jump in $\phi$ really an acceptable or approximate solution? 
  Let us place the surface at the local maximum of the potential
and linearize the equation of motion for $\phi$ around $\phi = 0$. Then
$\phi$ would be expected to role down to $-f_-$ on the inside and to
$f_+$ on the outside.  First we take a perturbative approach with $H(r)$
as a given background field, the same as above.  Defining $x=r-r_c$, and
considering
$|x|\ll r_c$, the equation just inside $(x<0)$ is
\begin{equation}
x\frac{d^2\phi}{dx^2} +\frac{d\phi}{dx} -\frac{1}{2l}\phi = 0
\end{equation}
and just outside $(x>0)$ is
\begin{equation}
x\frac{d^2\phi}{dx^2} +\frac{d\phi}{dx} +\frac{1}{l}\phi = 0 \, .
\end{equation}
The length scale, $l=1/\lambda f_+f_-r_c$, is very small compared to $r_c$
unless $f$ is comparable to the Planck mass.  The solutions to these
equations are
\begin{eqnarray}
\phi(x<0)&=& a_1 J_0\left(\sqrt{-2x/l}\right)
 + a_2 N_0\left(\sqrt{-2x/l}\right) \nonumber \\
\phi(x>0)&=& b_1 J_0\left(2\sqrt{x/l}\right)
 + b_2 N_0\left(2\sqrt{x/l}\right) \, .
\end{eqnarray}
If $p(x)$ were constant in the entire region, then 
the condition that $\phi(x=0)=0$ would mean that all coefficients
$a_1,a_2,b_1,b_2$ are zero.  
With the variation in $p(x)$ ignored, there is no smooth transition from
$-f_-$ on the inside to
$f_+$ on the outside, at least not in this perturbative approach.  
There must be singular behavior  at $x=0$.  After all, there is a horizon
separating the outside from the inside.

\begin{figure}[bth]\centering
\leavevmode 
\mbox{
\epsfxsize=8.5cm \epsfbox{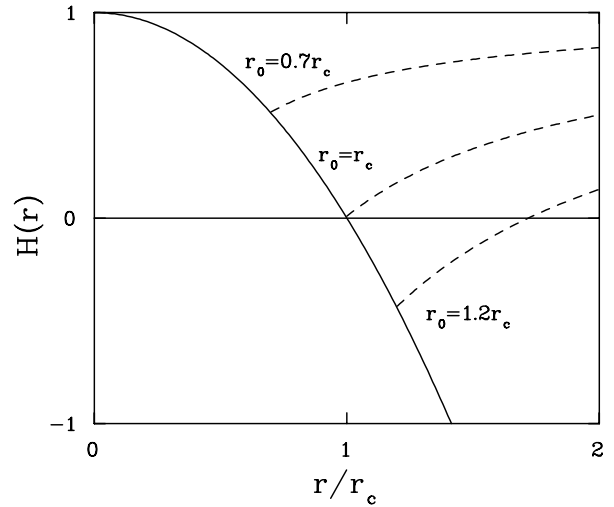}}
\caption{The metric function $H(r)$ resulting from the classical
field configuration $\phi(r)=-f_- \theta(r_0-r)+f_+ \theta(r-r_0)$.  There
is a horizon at every point where $H=0$.  $H(r)$ is given by the 
solid line for $r< r_0$ and by the dashed line for $r > r_0$.}
\label{H_plot}
\end{figure}

There is a simple way to understand this behavior.  
Eq.\ (\ref{Ein3}) in the surface region can be obtained by minimizing the
expression 
\begin{eqnarray}
\tilde I & = &  \int_{\rm surface} dx \, 
{1\over p(x)} \nonumber \\
& &  \times \bigg\{ \frac{1}{2} H(x)
\Big( \frac{d\phi}{dx}\Big)^2 + V\left(\phi(x)\right) \bigg\} \, .
\end{eqnarray}
Let $w$ characterize the width of the surface.  In order of magnitude
we can replace $d\phi/dx$ with $2f/w$, $V\left(\phi(x)\right)$ with
$\lambda f^4/4$, and $\int dx$ with $w$.  In the absence of gravity, $H =
p= 1$,   $\tilde I$ as a function of its
width is approximately
\begin{displaymath}
\frac{f^2}{w} + \lambda f^4 w \, .
\end{displaymath}
This is minimized when $w^2 \sim 1/\lambda f^2$, meaning that $w$ is
approximately equal to the correlation length, the usual result.  To
include gravity we replace $H(x)$ with $w/r_c$.  ($H$ vanishes linearly at
$r=r_c$ with negative slope on the inside and positive slope on the
outside.)   If $p(x)=1$,   $\tilde I$  as a function of
$w$ is approximately
\begin{displaymath}
\frac{f^2}{r_c} + \lambda f^4 w \, .
\end{displaymath}
This is minimized for $w=0$, forcing a jump in the field.   

The function $p(x)$, however,  varies from $p_{\rm in}$ to $p_{\rm out}$
at the surface as Eq.\ (\ref{Ein1}) dictates. 
Taking $\phi =  2fx/w$ in the surface region $-w/2 < x < w/2$, 
we have $p(x) = \exp (-\alpha r_c x/ w^2)$ where 
$\alpha = 16\pi G f^2 = 16\pi (f/m_P)^2$.
As $w\go 0$, $p_{\rm out} \go 0$
so that $\tilde I$ diverges.  Approximating  $H(x)$ by $-2x/r_c$
($x<0$) and $x/r_c$ ($x>0$) in the surface region, 
$\tilde I$ is given in terms of $y=w/(\alpha r_c)$ by
\beqn
&&\tilde I  = \alpha r_c \lambda f^4 F(y) ~, \cr
&&F(y) =\ \gamma [ g_1(y) + 2 g_1(-y)]  + [g_2(y) - g_2(-y)] ~,
\label{surface3}
\eeqn
where $\gamma  = 2\Delta f/9f$, 
$g_1(y) = y\{ y + ({1\over 2} -y) e^{1/2y} \}$, and
$g_2(y) = 8 y^4 (12 y^2 - 6y + 1) e^{1/2y}$.  The $g_1$ and $g_2$
represent contributions from the kinetic and potential terms,
respectively.   The constant $\alpha$ is tiny; it is $\sim 10^{-38}$
for $f=100$ MeV. $\gamma$ is smaller than  $ 10^{-2}$.  
$F(y) \sim y/6$ for   $y \gg 1$, whereas 
$F(y) \sim {1\over 2} \gamma y e^{1/2y}$ for $y \ll 1$.
For $\gamma < 10^{-2}$ the $g_2$-terms in $F(y)$ dominate except
for $y \ll 1$.  $\tilde I$ is  minimized at $y \sim 0.18$
almost independent of the value of $\gamma (< 10^{-2})$, or at 
$w \sim 4 (f/\lambda \Delta f)^{1/2} \cdot l_P$ where $l_P=1/m_P = 
1.6 \times 10^{-35}$m is the 
Planck length.  The surface is very thin compared with the size of
the bubble.  Simple calculations show that the surface energy is 
smaller than the volume energy by a factor of $f^3/(m_P^2 \Delta f)$.

The lifetime of the false vacuum may be determined semiclassically using the
methods of Coleman {\it et al.} without \cite{without} or with \cite{with}
gravity taken into account.  The rate per unit volume for making a
transition from the false vacuum to the true vacuum is expressed as
\begin{equation}
\frac{\Gamma}{V} = Ae^{-B/\hbar}[1+{\cal O}(\hbar)]
\end{equation}
where Planck's constant has been used here to emphasize the semiclassical
nature of the tunneling rate.  For the potential being used in this paper
we find the O(4), Euclidean space, bounce action (neglecting gravity) to be
\begin{equation}
B_0 = \frac{36\pi^2}{\lambda}\left(\frac{f}{\Delta f}\right)^3
\end{equation}
where the radius of the critical size bubble which nucleates the
transition is
\begin{equation}
\rho_c=\frac{3}{\Delta f} \sqrt{\frac{2}{\lambda}} \, .
\end{equation}
For any sensible estimate of the coefficient $A$ the lifetime of the false
vacuum will exceed the present age of the universe when the condition
\begin{equation}
\lambda \left(\frac{\Delta f}{f}\right)^3 < 1
\end{equation}
is satisfied.  This calculation is based on the thin wall approximation,
which  is valid when the critical radius is large compared to the coherence
length of  the potential, namely $1/\sqrt{|V^{\prime\prime}|}$.  This
condition translates  into
$\Delta f \ll 6f$.
With gravity included the bounce action is
\begin{equation}
B = \frac{B_0}{\left[1+(\rho_c/2r_c)^2 \right]^2} \, .
\end{equation}
Gravitational effects are negligible when $\rho_c \ll r_c$, 
or $f \sqrt{f/\Delta f}  \ll m_P$.  When $\rho_c > r_c$,   the black hole
is too small to accommodate even a single  nucleation bubble, or in other
words the bounce solution does not have O(4) symmetry.
Nucleation of a bubble is further suppressed.

The proper time for a light signal  starting  a distance
$\delta \ll r_c$ from the event horizon to propagate to the origin, 
as computed with the de Sitter metric,  is $\Delta t= {1\over 2} r_c
\ln (2 r_c/\delta)$.  For 
$\delta$ of order of the surface thickness this is approximately
$r_c \ln (m_P/f)$, which is much bigger than $r_c$ for $f \ll m_P$.
This is the minimum time for a symmetry restoring signal,  originating
near the surface, to destroy the bubble.

An outside observer would have no information on the state of the universe
inside the Schwarzschild radius.  Such an observer would see a black hole
with mass $M_c = m_P^2 r_c/2$.  An observer on the inside, however, would be
living in a de Sitter space with a cosmological constant $\Lambda =
3/r_c^2$.
A standard change of coordinates \cite{Weinberg} puts the
metric on the inside in the form
\begin{equation}
d\tau^2 = dt'^2 -e^{2t'/r_c}\left(dr'^2 + r'^2 d\Omega'^2\right) \, .
\end{equation}
From these considerations we may identify a Hubble constant = $1/r_c$.

To make matters interesting, let us suppose that the cosmological
constant suggested by recent observations of distant Type Ia supernovae
\cite{super} arises from the universe actually being in a false vacuum
state.  A best fit to all cosmological data \cite{science} reveals that
the present energy density of the universe has the critical value of
$\epsilon_c = 3H_0^2/8\pi G$, with one-third of it consisting of
ordinary matter and two-thirds of it contributed by the cosmological
constant.  With a present value of the Hubble constant of $H_0=65$
km/s$\cdot$Mpc we find that
\begin{equation}
r_c = \sqrt{\frac{3}{2}}\frac{1}{H_0} = 1.7 \times 10^{26} \,\, {\rm m}
= 5.5 \, {\rm Gpc}\end{equation}
and 
\begin{equation}
\left( \lambda f^3 \Delta f \right)^{1/4} = \epsilon_c^{1/4}
=2.4 \times 10^{-3} \,\, {\rm eV} 
\, .
\end{equation}
This constraint on the parameters of the potential, $\lambda$, $f$ and
$\Delta f$, is entirely consistent with the constraints imposed by the
lifetime of the false vacuum exceeding the present age of the universe.

The above discussion is pure speculation of course.  A cosmological
constant, if it exists, may have its origins elsewhere.  But if it does
arise from a false vacuum, a variety of questions immediately present
themselves.  Is $\phi$ a new field, not present in the standard model of
particle physics, whose only purpose is this?  Where does the energy
scale of 2.4  meV  come from?  Why should
$V(\phi)$ have a global minimum of 0, especially when quantum mechanical
fluctuations are taken into account?  To these questions we have no
answers.

Another amusing possibility is that black holes believed to exist near
the center of galaxies are false vacuum black holes.  The mass of
the critical false vacuum black hole is given by
\beeq
M_c = \sqrt{ {3\over 32\pi G^3 \epsilon_-} }
= {0.282 \cdot M_{\rm Sun} \over
\sqrt{ \epsilon_-/{\rm GeV}^4}} 
\label{critical1}
\eneq
where $M_{\rm Sun}$ is the solar mass.  In other words the scale
${\epsilon_-}^{1/4} = 1$ MeV corresponds to a black hole of 
1 million solar mass.

Yet another amusing possibility is an analogy to spin glasses \cite{glass}
and, possibly, QCD.  A spin glass is characterized by a phase space that
has a complicated landscape of troughs and valleys.  At low temperatures
barriers become very large, and the system may become trapped in a
metastable state for the whole duration of an experiment.
The possibility of a second minimum in the effective potential has been
noted in regards to the chiral phase transition in QCD \cite{proximal}. 
Chiral symmetry is explicitly broken by the nonzero masses of the up and
down quarks.  In the context of a low energy representation of QCD by the
linear sigma model, it was argued in \cite{proximal} that one should
consider the most general symmetry breaking potential which is at most
fourth order in the fields.
\begin{equation}
V_{SB} \,=\, - \sum_{n=1}^4 \frac{\delta_n}{n!}\sigma^n
\,+\, (\delta_5\sigma + \delta_6 \sigma^2)
\mbox{\boldmath $\pi$}^2 \,.
\end{equation}
(Other symmetry-breaking terms one might think of adding simply amount to a
redefinition of the eight parameters $\lambda, f, \delta_n$.) 
Historically the most used potential is $-f_{\pi} m_{\pi}^2 \sigma$.  The
full potential then has a global minimum at $\sigma = f_{\pi}$ and a saddle
point near $\sigma = - f_{\pi}$ (it is a local minimum in the $\sigma$
direction but a local maximum in the $\mbox{\boldmath $\pi$}$ direction). 
Another one is $\frac{1}{2} m_\pi^2
\mbox{\boldmath $\pi$}^2$, which has two degenerate minima at $\sigma=\pm
f_{\pi}$.  It is possible to choose the symmetry breaking potential so
that the full potential has two nearly degenerate minima in the full
$\sigma - \mbox{\boldmath $\pi$}$ space such that small field fluctuations
around one or the other of them reproduces known physics.  A natural
estimate is $\lambda = 10$, $f = 100$ MeV, and $\Delta f = 2$ MeV.  Then
$r_c = 230$ km,  and such a black hole would have a mass of about 80 suns. 
Such astronomical objects may be left over from the QCD phase transition in
the early universe.  One may even  speculate that they formed the seeds for
galactic black holes.

The most outrageous possibility is that we might be living inside an
enormous  black hole.  If so, what is outside our universe?

\section*{Acknowledgements}

This work was supported in part by the US Department of Energy under
grants DE-FG02-87ER40328 and DE-FG02-94ER-40823.




\end{document}